\documentclass[twocolumn,showpacs,preprintnumbers]{revtex4}
\usepackage{amsmath}
\usepackage{graphicx}
\usepackage{dcolumn}
\usepackage{bm}

\begin{document}

\title{ Generalized quantum Rabi model with both one- and two-photon terms:
A concise analytical study}
\author{You-Fei Xie$^{1}$, Liwei Duan$^{1}$, and Qing-Hu Chen$^{1,2,*}$}

\address{
$^{1}$ Department of Physics, Zhejiang University, Hangzhou 310027,
 China \\
$^{2}$  Collaborative Innovation Center of Advanced Microstructures,  Nanjing University,  Nanjing 210093, China
 }\date{\today }

\begin{abstract}
A generalized quantum Rabi Hamiltonian with both one- and two-photon terms
has emerged in the circuit quantum electrodynamics system for a decade. The
usual parity symmetry is broken naturally in the simultaneous presence of
both couplings, which complicates analytical treatments, even in the
rotating wave approximations. In this paper, we propose an adiabatic
approximation to this generic model by using Bogoliubov operators, and
obtain a very concise analytical solution for both eigenvalues and
eigenstates. Although the adiabatic approximation is only exact in the
vanishing limit of the qubit frequency, the results for some physical
observables nevertheless agree well with the numerical ones in a wide
parameter regime. In the rotating-wave approximations, we also derive an
analytical eigensolution. Two dominant Rabi frequencies are found in the
Rabi oscillations of this generalized model. We also apply the present
analytical theory to the vacuum Rabi splitting. It is found that some new
phenomena emerge just because of the presence of the additional two-photon
coupling term.
\end{abstract}

\pacs{03.65.Yz, 03.65.Ud, 71.27.+a, 71.38.k}
\maketitle

\section{Introduction}

The quantum Rabi model (QRM) describes the most simple and at the same time
the most important coupling between a continuous degree of freedom (a mode
of the light field) and a discrete one (a two-level system or qubit) which
is linear in the quadrature operators~\cite{Braak2}. In addition, the
nonlinear coupling appears naturally as an effective model for a three-level
system when the third (off-resonant) state can be eliminated. The two-photon
model has been proposed to apply to certain Rydberg atoms in superconducting
microwave cavities~\cite{Bertet,Brune}. Recently, a realistic implementation
of the two-photon QRM using trapped ions has been proposed~\cite{Felicetti}.
Usually, the two-photon term is the secondary effect, and thus limited to
the weak-to-moderate coupling regime typical for experimental setups within
cavity or circuit quantum electrodynamics (QED) systems .

Here we study a natural generalization of the QRM which exhibits both linear
\ and non-linear couplings between both constituents, i.e. the QRM with both
one- and two-photon terms, with Hamiltonian
\begin{equation}
H=\frac{\Delta }{2}\sigma _{z}+\omega a^{\dagger }a+\sigma _{x}\left\{
g_{1}\left( a^{\dagger }+a\right) +g_{2}\left[ \left( a^{\dagger }\right)
^{2}+a^{2}\right] \right\} ,  \label{12p-rabimodel}
\end{equation}%
where $\Delta $\ and $\omega $ are respectively frequencies of qubit and
cavity, $\sigma _{x,z}$ are Pauli matrices describing the two-level system, $%
a$ ($a^{\dagger }$) are the annihilation (creation) bosonic operators of the
cavity mode, and $g_{1}$ ($g_{2}$) is the one-photon (two-photon)
qubit-cavity coupling constant.

The generalized QRM ~\cite{Ying} described by Eq. (\ref{12p-rabimodel}) has
actually been realized in a flux qubit coupled to the plasma mode of its
superconducting quantum interference device (SQUID) detector to reduce the
dephasing rate ~\cite{Bertet2,Bertet3} in 2005. Both coupling strengths $%
g_{1}$ and $g_{2}$ depend on the SQUID biased current. Recently, Felicetti
\textsl{et al.} proposed a different circuit QED setup where a dc-SQUID is
inductively coupled to a flux qubit and a current bias is added to the SQUID
~\cite{Felicetti1}. Expanding the qubit-SQUID interaction up to the second
order can yield both linear and nonlinear interaction terms, although their
original intention is to implement the two-photon QRM by setting zero dc
current biases. Most recently, Pedernales \textsl{et al.} suggested that a
background of a ($1+1$)-dimensional black hole requires a QRM with one- and
two-photon terms that can be implemented in a trapped ion for the quantum
simulation of Dirac particles in curved spacetime ~\cite{Pedernales}. So the
QRM with both one- and two-photon couplings not only is a generic model in
the circuit QED, but also has wide applications in the area of
cross-disciplinary research.

Both the one-photon QRM \cite{Braak2} and two-photon QRM \cite{Ng} have been
studied extensively for a few decades (for a review, please refer to Refs. ~%
\cite{Lee,yuxi,ReviewF}). The analytical exact solutions based on the
well-defined transcendental function have been only found recently for
one-photon ~\cite{Braak} and two-photon~\cite{Chen2012} QRM. These solutions
have stimulated extensive research interests in the exact solutions to the
QRM with both one-photon term~\cite%
{Zhong,Maciejewski,Chilingaryan,Peng,Wang, Fanheng} and two-photon term ~%
\cite{Trav,Maciejewski2,duan2016,Zhangyz}. Many analytical approximate but
still very accurate results have been also given ~\cite%
{Feranchuk,Irish,chenqh,zheng,chen2,luo2,zhang,Zhiguo}. In some model
parameter limits, the dynamics and quantum criticality have been also
studied exactly ~\cite{Casanova,plenio,hgluo}.

In the QRM with both linear and nonlinear couplings, the parity symmetry is
broken naturally, and the analytical solution becomes more difficult, in
contrast to the unmixed models. To the best of our knowledge, no analytical
solutions to these mixed model are available to date. In this paper, we
propose concise analytical solutions to this generic model, which may be
used as a solid basis for further advanced analytical studies. On the other
hand, it could be easily applied to the circuit QED experiments where the
expansion of qubit-oscillator interaction in the second-order is not
negligible. As a preliminary application, we study effects of the nonlinear
coupling on the vacuum Rabi splitting~\cite{sanche} by using these
analytical solutions.

The paper is structured as follows: In Sec. II, by Bogoliubov operators, we
propose a general scheme to solve the QRM with both one- and two-photon
couplings. In Sec. III, the adiabatic approximations are made and concise
analytical solutions are derived. The analytical energy level and qubit
population dynamics are then presented and analyzed in Sec. IV. Section V is
devoted to an analytic analysis in the rotating-wave approximations (RWAs).
The applications of the analytical theories both in the adiabatic
approximations and the RWA to the vacuum Rabi splitting are given in Sec.
VI. The last section contains some concluding remarks.

\section{ Operator transformations and numerically exact scheme}

Associated with unmixed QRMs are the conserved parity $\Pi _{1p}=-\sigma
_{z}\exp \left( i\pi a^{+}a\right) $ for one-photon \cite{Braak2} and $\Pi
_{2p}=-\sigma _{z}\exp \left( \frac{i\pi }{2}a^{+}a\right) $ for two-photon
\cite{duan2016} qubit-cavity interaction, such that $\left[ H,\Pi \right] =0$%
. $\Pi $ has two eigenvalues ($\pm 1$) for the one-photon QRM, while it has
four eigenvalues ($\pm 1,\pm i$) for the two-photon QRM, because one can
partition the Hilbert space in two subspaces, each with an SU(1,1) symmetry
for the field depending on the photonic number. So in the two unmixed
models, the parity symmetry acting in the bosonic Hilbert space greatly
facilitates the study even in the analytical analysis. However, in the
present generalized model with both one- and two-photon interactions with
the qubit, no such conserved parity is available, so the analytical study is
still challenging.

In this section, we introduce a scheme to pave the way to the solutions. For
convenience, we write a transformed Hamiltonian with a rotation around the $%
y $ axis by an angle $\frac{\pi }{2}$ in the matrix form in units of $\omega
=1,$%
\begin{widetext}
\begin{equation}
H=\left(
\begin{array}{ll}
a^{\dagger }a+g_{1}\left( a^{\dagger }+a\right) +g_{2}\left[ \left(
a^{\dagger }\right) ^{2}+a^{2}\right] & ~~~~~~~~-\frac{\Delta }{2} \\
~~~~~~~~-\frac{\Delta }{2} & a^{\dagger }a-g_{1}\left( a^{\dagger }+a\right)
-g_{2}\left[ \left( a^{\dagger }\right) ^{2}+a^{2}\right]%
\end{array}%
\right). \label{H2}
\end{equation}
\end{widetext}

To obtain a clean manifold, we perform Bogoliubov transformation to the
bosonic degree of freedom so that the first diagonal element in Hamiltonian
matrix (\ref{H2}) only consists of a free transformed bosonic number
operator. The new bosonic operator is introduced as
\begin{equation}
A=ua+va^{\dagger }+w,A^{\dagger }=ua^{\dagger }+va+w,
\end{equation}%
while the corresponding number state would be
\begin{equation}
\left\vert n\right\rangle _{A}=S(r)D^{\dag }(w)\left\vert n\right\rangle ,
\end{equation}%
where $\left\vert n\right\rangle $ is the number state in original Fock
space, $r=arc\cosh u$, $S(r)$ is the squeezing operator, and $D(w)$ is the
displaced operator
\begin{equation*}
S(r)=e^{\frac{r}{2}(a^{2}-a^{\dag 2})},D(w)=e^{w(a^{\dag }-a)}.
\end{equation*}%
If set
\begin{equation}
u=\sqrt{\frac{1+\beta }{2\beta }},\;v=\sqrt{\frac{1-\beta }{2\beta }},w=%
\frac{u^{2}+v^{2}}{u+v}g_{1},
\end{equation}%
with $\beta =\sqrt{1-4g_{2}^{2}}$, we have a simple quadratic form of the
first diagonal element as
\begin{equation*}
H_{11}=\frac{A^{\dagger }A-v^{2}-w^{2}}{u^{2}+v^{2}},
\end{equation*}%
which eigenstates are just $\left\vert n\right\rangle _{A}$.

Similarly, we can introduce another operator $B$,
\begin{equation}
B=ua-va^{\dagger }+w^{\prime },B^{\dagger }=ua^{\dagger }-va+w^{\prime },
\end{equation}%
with the corresponding number state
\begin{equation}
\left\vert n\right\rangle _{B}=S^{\dag }(r)D^{\dag }(w^{\prime })\left\vert
n\right\rangle .
\end{equation}%
A simple quadratic form of the second diagonal element would be achieved if
set $w^{\prime }=\frac{u^{2}+v^{2}}{v-u}g_{1}$:
\begin{equation*}
H_{22}=\frac{B^{\dagger }B-v^{2}-w^{\prime 2}}{u^{2}+v^{2}}.
\end{equation*}

In terms of the Bogoliubov operators $A$ and $B$, the Hamiltonian can be
written as
\begin{equation}
H=\left(
\begin{array}{ll}
\beta \left( A^{\dagger }A-v^{2}-w^{2}\right) & ~\ \ \ \ -\frac{\Delta }{2}
\\
~~\ \ -\frac{\Delta }{2} & \;\;\;\beta \left( B^{\dagger }B-v^{2}-w^{\prime
2}\right)%
\end{array}%
\right) .
\end{equation}%
Because now only the number operators $A^{+}A$ and $B^{+}B$ are present, in
principle, the wavefunction can be expanded in terms of the number states of
these new operators as
\begin{equation}
\left\vert {}\right\rangle =\left(
\begin{array}{l}
\sum_{n=0}c_{n}\left\vert n\right\rangle _{A} \\
\sum_{n=0}d_{n}\left\vert n\right\rangle _{B}%
\end{array}%
\right) .  \label{wavefunction}
\end{equation}

Note that here the Hilbert space can be decomposed into different $n$
manifolds spanned by the spin and oscillator basis of $\left\vert \uparrow
\right\rangle \left\vert n\right\rangle _{A}$and$\ \left\vert \downarrow
\right\rangle \left\vert n\right\rangle _{B}$, where $\left\vert \uparrow
\right\rangle \ \left( \left\vert \downarrow \right\rangle \right) $ denotes
the upper (lower) states of the qubit. Inserting Eq. (\ref{wavefunction})
into the Schr$\overset{..}{o}$dinger equation, we have
\begin{eqnarray}
\beta \left( m-v^{2}-w^{2}\right) c_{m}-\frac{\Delta }{2}\sum_{n=0\
}^{N_{tr}}D_{mn}d_{n} &=&Ec_{m},  \label{exact1} \\
\beta \left( m-v^{2}-w^{\prime 2}\right) d_{m}-\frac{\Delta }{2}%
\sum_{n=0}^{N_{tr}}D_{nm}c_{n} &=&Ed_{m},  \label{exact2}
\end{eqnarray}%
where $N_{tr}$ is the truncation number,
\begin{widetext}
\begin{eqnarray*}
D_{mn} =\sqrt{n!m!\beta }\left( uv\beta \right) ^{(m+n)/2}\exp \left( \frac{%
-2g_{1}^{2}}{\beta ^{3}}\right) \sum_{i=0}^{\min (m,n)}
\frac{(-1)^{(m-i)/2}\left( uv\right) ^{-i}}{i!(m-i)!(n-i)!}H_{m-i}\left(
\frac{g_{1}(u-v)}{\beta ^{3/2}\sqrt{-uv}}\right) H_{n-i}\left( \frac{%
-g_{1}(u+v)}{\beta ^{3/2}\sqrt{uv}}\right).
\end{eqnarray*}%
\end{widetext}
Here $H_{n}(x)$ stands for the Hermite polynomials.

In principle, based on Eqs. (\ref{exact1}) and (\ref{exact2}), we can obtain
exactly the spectra of the generalized QRM to any desired accuracy by
increasing the truncation number of the summation. For the numerical data
presented below, we typically select $N_{tr}=60$, and converging results
with relative errors less than $10^{-6}$ are convincingly arrived at.

\section{Adiabatic approximations}

In this section, we turn to an analytical scheme. In the framework of Eqs. (%
\ref{exact1}) and (\ref{exact2}), analytical approximations can be performed
systematically. As a zero-order approximation, we only consider transition
between states in the same manifold spanned by $\left\vert \uparrow
\right\rangle \left\vert m\right\rangle _{A}$and$\ \left\vert \downarrow
\right\rangle \left\vert m\right\rangle _{B}$, then we have%
\begin{eqnarray*}
\beta \left( m-v^{2}-w^{2}\right) c_{m}-\frac{\Delta }{2}d_{m}D_{mm}
&=&Ec_{m}, \\
\beta \left( m-v^{2}-w^{\prime 2}\right) d_{m}-\frac{\Delta }{2}c_{m}D_{mm}
&=&Ed_{m}.
\end{eqnarray*}%
Nonzero coefficients give the following equation
\begin{eqnarray*}
&&\left[ E-\beta \left( m-v^{2}-w^{2}\right) \right] \left[ E-\beta \left(
m-v^{2}-w^{\prime 2}\right) \right] \\
&&-\frac{\Delta ^{2}}{4}D_{mm}^{2}=0.
\end{eqnarray*}%
The eigenvalues are then easily given by
\begin{eqnarray}
E_{m}^{\pm } &=&\beta \left( m-v^{2}\right) -\frac{g_{1}^{2}}{\beta ^{2}}
\notag \\
&&\pm \frac{1}{2}\sqrt{\beta ^{2}(w^{2}-w^{\prime 2})^{2}+\Delta
^{2}D_{mm}^{2}}.  \label{Full_en}
\end{eqnarray}%
The corresponding eigenstate is
\begin{equation}
\left\vert m\right\rangle _{\pm }\propto \left(
\begin{array}{l}
\ \ \ \ \ \ \ \frac{1}{2}\Delta D_{mm}\left\vert m\right\rangle _{A} \\
\left[ \beta (m-v^{2}-w^{2})-E_{m}^{\pm }\right] \left\vert m\right\rangle
_{B}%
\end{array}%
\right) .  \label{Full_w}
\end{equation}

As noted from Eqs. (\ref{exact1}) and (\ref{exact2}), it is just the qubit
frequency $\Delta $ that correlates the different manifolds. The zero-order
approximate results Eqs. (\ref{Full_en}) and (\ref{Full_w}) should be exact
for the model if the qubit frequency vanishes, because the transitions among
any manifolds disappear automatically. In literature, this approximation is
also called the adiabatic approximation \cite{adi}, which works best for
small qubit frequency and strong coupling. For large qubit frequency, and
weak coupling, the high order approximation should be performed. Actually,
we can further consider the transitions between more neighboring manifolds,
and more complicated analytical results would be derived. In the recent
circuit QED, since the qubit frequency is usually smaller than the frequency
of oscillator and the oscillator-qubit interaction has entered the
ultrastrong \cite{Niemczyk,exp,tiefu}, even deep-strong-coupling regime \cite%
{Yoshihara}, the adiabatic approximation should work well. The further
high-order corrections are not discussed here, and left for future study.

In the applications to the realistic circuit QED system, we also need to
consider the full qubit Hamiltonian $H=-\left( \epsilon \sigma _{z}+\Delta
\sigma _{x}\right) /2,$ where $\epsilon $ is the static bias. The
eigenenergy can be easily derived as
\begin{eqnarray}
E_{m}^{\pm }=&&\beta \left( m-v^{2}\right) -\frac{g_{1}^{2}}{\beta ^{2}}
\notag \\
&&\pm \frac{1}{2}\sqrt{(\beta w^{2}-\beta w^{\prime 2}+\epsilon )^{2}+\Delta
^{2}D_{mm}^{2}}.  \label{En-bias}
\end{eqnarray}%
\

\section{Energy levels and dynamics}

With the recent progress in technology, the one-photon coupling term has
reached the ultrastrong-coupling regime ($g_{1}/\omega \approx 0.12$)
experimentally with superconducting flux qubits inductively coupled to
superconducting resonators \cite{Niemczyk,exp,tiefu}. More recently, it has
accessed to the deep-strong-coupling regime ($g_{1}/\omega \approx 1.34$)
\cite{Yoshihara,Forn2}. The two-photon term emerges from the second process
in cavity QED or the expansions of the qubit-oscillator interaction up to
the second order in the circuit QED, so the two-photon terms should not be
strong generally. Physically, its dimensionless coupling strength $%
g_{2}/\omega $ should be less than the interaction-induced spectral collapse
point $0.5$ \cite{Ng,Felicetti,duan2016}.

To show the validity of the adiabatic approximation to the recent
experimentally accessible systems, we examine some physical observables such
as the energy spectra and dynamics for the one-photon coupling strength $%
g_{1}$ ranging from weak to deep strong coupling, while the two-photon
coupling strength $g_{2}$ is fixed to be a moderate value. The energy
spectra can directly account for the experimental transmission spectrum, and
dynamics can be measured experimentally.

\begin{figure}[tbp]
\includegraphics[width=8cm]{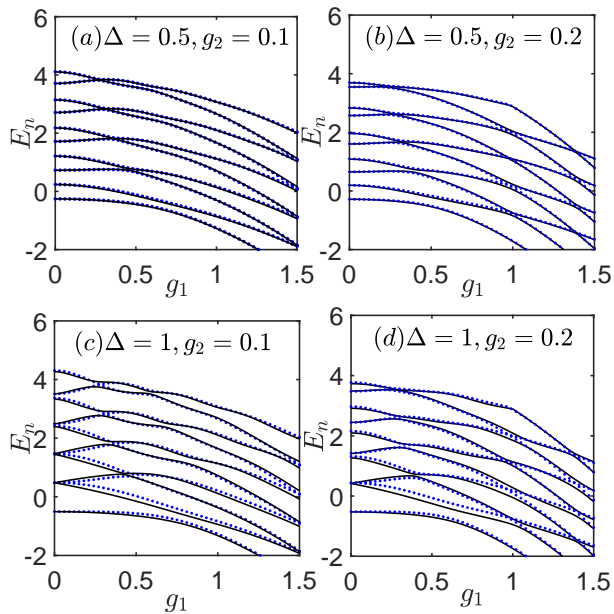}
\caption{ (Color online) The energy levels (blue filled circles) as a
function of $g_1$ for $\Delta =0.5$, $g_{2}=0.1$ and $0.2$ (upper panel) and
for $\Delta =1$, $g_{2}=0.1$ and $0.2$ (lower panel). Numerical exact
results are indicated by black solid lines.}
\label{energy_full}
\end{figure}

The energy levels by Eq. (\ref{Full_en}) as a function of $g_{1}$ for $%
g_{2}=0.05$ and $0.1$, at the qubit frequency $\Delta =0.5$ and $1$ are
displayed in Fig. \ref{energy_full}. The numerically exact results by using
Eqs. (\ref{exact1}) and (\ref{exact2}) are also collected. By the way, the
exact results can be also obtained by the numerically diagonalization in
original Fock space. For $\Delta =0.5$, the present results are in excellent
agreement with the exact ones, while for $\Delta =1$ the adiabatic
approximation still gives the good results. Actually, the present approach
is basically a perturbation in the qubit frequency $\Delta ,$ so with
increasing $\Delta $, the present results show a little bit deviation from
the true spectra. Practically, in the recent experiments, $\Delta /\omega $
is usually not larger than $1\ $\cite{Niemczyk,exp,tiefu,Yoshihara}. \ In
the table I of Ref. ~\cite{Yoshihara}, the value of $\Delta /\omega $ is
even in the order of magnitude of $0.01$. In this sense, our results should
be convincingly applicable to these circuit QED experiments if the
two-photon coupling effect could be involved.
\begin{figure}[tbp]
\includegraphics[width=8cm]{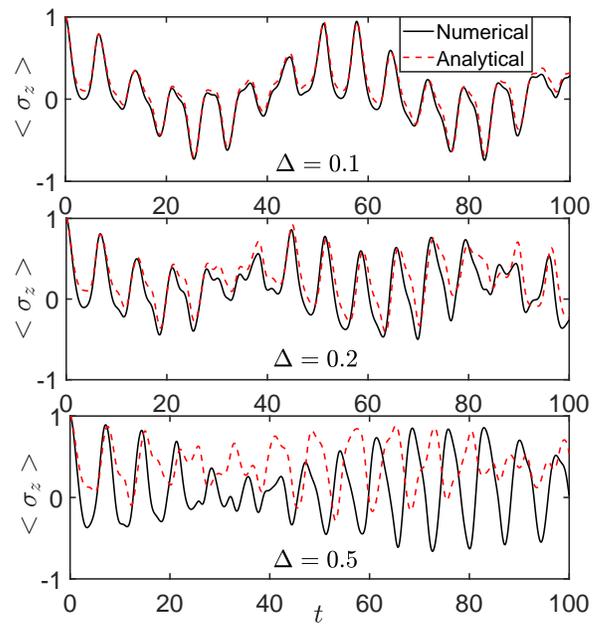}
\caption{ (Color online) Population difference at $\Delta =0.1,0.2$ and $0.5$
(from top to bottom) for $g_{1}=0.5,g_{2}=0.1$ initiated from photonic
vacuum and upper level, which are denoted by red dashed lines. The numerical
results are denoted by black solid lines. }
\label{diff_full}
\end{figure}

We then examine the dynamics of the population difference $\langle \sigma
_{z}(t)\rangle $ for different qubit frequencies $\Delta =0.1,0.2$, and $0.5$
and the coupling strength $g_{1}=0.5,g_{2}=0.1$ in Fig. \ref{diff_full}. For
small value of $\Delta $, the adiabatic approximation can describe the
dynamics almost exactly in the long time scale. For large $\Delta $, our
analytical results can match the oscillation in phase for a long time. It
only begins to get out of phase after many periods of oscillations. For all
cases studied, our theory can basically give right description of the
dynamics. Note that the parameters ($\Delta $ and $g_{1}$) we used here are
actually included but not limited to those in the recent experiments ~\cite%
{Niemczyk,exp,tiefu,Yoshihara}.

\section{Analytical analysis in the Rotating-wave approximations}

In the RWA, the one-photon QRM is reduced to Jaynes-Cummings model~\cite{JC}%
, which always reveals the general aspects of the full model at the weak
coupling, such as Rabi oscillations, collapses and revivals of quantum state
populations, entanglement dynamics, Schr$\overset{..}{o}$dinger cat states.

Similar to the standard unmixed QRMs, the RWA can be made for the mixed
model by neglecting the counter rotating terms, $g_{1}\left( a^{\dagger
}\sigma _{+}+a\sigma _{-}\right) ,g_{2}\left[ \left( a^{\dagger }\right)
^{2}\sigma _{+}+a^{2}\sigma _{-}\right] $, which gives the Hamiltonian in
the RWA as
\begin{equation}
H=a^{\dagger }a+\frac{\Delta }{2}\sigma _{z}+g_{1}\left( a^{\dagger }\sigma
_{-}+a\sigma _{+}\right) +g_{2}\left[ \left( a^{\dagger }\right) ^{2}\sigma
_{-}+a^{2}\sigma _{+}\right] .  \label{H_RWA}
\end{equation}%
It takes the following matrix form in the basis of $\sigma _{z}$%
\begin{equation}
H=\left(
\begin{array}{ll}
\ \ a^{\dagger }a+\frac{\Delta }{2} & g_{1}a+g_{2}a^{2} \\
g_{1}a^{\dagger }+g_{2}\left( a^{\dagger }\right) ^{2} & \ a^{\dagger }a-%
\frac{\Delta }{2}%
\end{array}%
\right) .
\end{equation}

For one-photon case, the energy level for the QRM in the RWA reads \cite%
{Scully}
\begin{equation*}
E_{n,1p}^{(k)}=n+\frac{1}{2}+\left( -1\right) ^{k}\frac{1}{2}\sqrt{\left(
\Delta -1\right) ^{2}+4g_{1}^{2}\left( n+1\right) },k=1,2.
\end{equation*}%
with eigenfunctions consisting of $\left\vert n\right\rangle \left\vert
\uparrow \right\rangle $ and $\left\vert n+1\right\rangle \left\vert
\downarrow \right\rangle $. The RWA result for the eigenenergy for
two-photon QRM \cite{Albert} is given by

\begin{equation*}
E_{n,2p}^{(k)}=n+1+\left( -1\right) ^{k}\frac{1}{2}\sqrt{\left( \Delta
-2\right) ^{2}+4g_{2}^{2}\left( n+2\right) \left( n+1\right) },
\end{equation*}%
with eigenfunctions consisting of $\left\vert n\right\rangle \left\vert
\uparrow \right\rangle $ and $\left\vert n+2\right\rangle \left\vert
\downarrow \right\rangle $.

For the case with both one- and two-photon couplings, we first propose the
wavefunction as
\begin{equation}
\left\vert n\right\rangle _{1}=\left( \
\begin{array}{l}
\;\;\;c_{n}\left\vert n\right\rangle \\
e_{n}\left\vert n+1\right\rangle +f_{n}\left\vert n+2\right\rangle%
\end{array}%
\right) ,n=0,1,2,...  \label{wave1}
\end{equation}%
which includes the basic process of the two unmixed models. One should note
that the Hamiltonian (\ref{H_RWA}) cannot be decomposed cleanly into
independent $n\ $sub-space $R_{n}=\{\left\vert n\right\rangle \left\vert
\uparrow \right\rangle ,\left\vert n+1\right\rangle \left\vert \downarrow
\right\rangle ,\left\vert n+2\right\rangle \left\vert \downarrow
\right\rangle \}$, because the interaction also couples the states in
different subspaces, in sharp contrast to both unmixed models. So the true
wavefunction should include the contribution from infinite bare states in
all subspaces. This is to say, unlike the Jaynes-Cummings model, the mixed
model in the RWA is not easy to solve. But it is still expected that the
wavefunction (\ref{wave1}) defined in the $n\ $sub-space $R_{n}$ would be
very accurate because dominant processes have been described.

By the Schr$\overset{..}{o}$dinger equation, we can get an univariate cubic
equation, which is given in detail in the appendix. Generally, there are
three different real roots as listed in the end of the appendix. We observed
that only the first root $\lambda _{1}$ from Eq. (\ref{y1}) is closest to
the numerical ones, which is denoted by $E_{n}^{(1)}$ for the generalized
model. One can easily find that $E_{n}^{(1)}$ for $g_{2}=0$\ ($g_{1}=0$)
here is reduced to $E_{n,1p}^{(1)}$ $\left( E_{n,2p}^{(1)}\right) $.

Both processes in the unmixed models can be also incorporated in another
wavefunction
\begin{equation}
\left\vert n\right\rangle _{2}=\left( \
\begin{array}{l}
f_{n}^{\prime }\left\vert n-1\right\rangle +c_{n}^{\prime }\left\vert
n\right\rangle \\
\ \ \ e_{n}^{\prime }\left\vert n+1\right\rangle%
\end{array}%
\right) ,\ n=0,1,2,...  \label{wave2}
\end{equation}%
Similarly, we can obtain another univariate cubic equation, which is also
given in Appendix A. We find that only the third root $\lambda _{3}$ in Eq. (%
\ref{y3}) is closest to the numerical ones, which is denoted by $E_{n}^{(2)}$
for the generalized model. Analogously, $E_{n}^{(2)}$ for $g_{2}=0$\ ($%
g_{1}=0$) here is simplified to $E_{n,1p}^{(2)}$ $\left(
E_{n,2p}^{(2)}\right) $.

\begin{figure}[tbp]
\includegraphics[width=8cm]{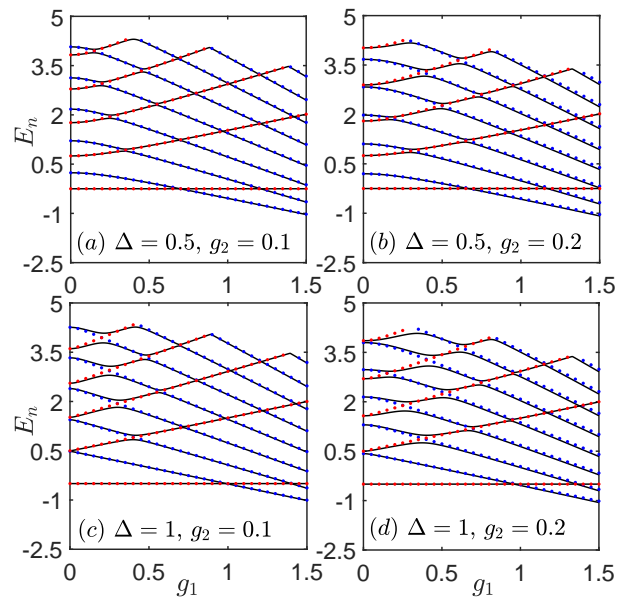} %
\caption{(Color online) The energy levels as a function of $g_{1}$ for fixed
$g_{2}$ for different $\Delta =0.5$ (upper panel) and $1$ (lower panel). The
numerical results are denoted by black solid lines. Blue and red dotted
lines denote $E_{n}^{(1)}$ and $E_{n}^{(2)}$ respectively.}
\label{energy_RWA}
\end{figure}

To show the validity of the approach with a few bare states, we compare the
analytical results for the energy spectrum with the numerical ones. We also
study the effect of moderate two-photon terms similar to the last section.
The energy spectra as a function of $g_{1}$ with fixed $g_{2}=0.1,0.2$ for $%
\Delta =0.5$ and $\ 1.0$ are displayed in Fig. \ref{energy_RWA}. It is
interesting to note that the analytical results agree well with the
numerical ones. Our result based on a few Fock states deviates from the
exact ones slightly around the avoided crossings. In other words, around the
avoided crossings, the contributions from the neighboring sub-spaces should
be considered. It should be pointed out that the lowest red energy level in
Fig. \ref{energy_RWA} can not be brought into the above general expression.
Its energy is $-1/2$ with photonic vacuum states in atomic lower level.

\textsl{Rabi oscillations.-} With the eigensolutions obtained above, we can
recheck many remarkable facets observed in the one-photon QRM in the RWA.
Here we study the interplay effect of both couplings on the celebrated Rabi
oscillation. Initiated from the number state in the upper level $\left\vert
t=0\right\rangle =\left\vert n\right\rangle \left\vert \uparrow
\right\rangle $, which can be expanded in terms of normalized eigenstates
Eqs. (\ref{wave1}) and (\ref{wave2})%
\begin{equation*}
\left\vert n\right\rangle \left\vert \uparrow \right\rangle =d_{1}\left\vert
n\right\rangle _{1}+d_{2}\left\vert n\right\rangle _{2}+d_{3}\left\vert
n+1\right\rangle _{2},
\end{equation*}%
where
\begin{equation*}
d_{1}=c_{n},d_{2}=c_{n}^{\prime },d_{3}=f_{n+1}^{\prime },
\end{equation*}%
the time-dependent wavefunction is then determined as
\begin{eqnarray*}
&&\left\vert \psi (t)\right\rangle =c_{n}e^{-iE_{n}^{(1)}t}\left\vert
n\right\rangle _{1}+c_{n}^{\prime }e^{-iE_{n}^{(2)}t}\left\vert
n\right\rangle _{2} \\
&&+f_{n+1}^{\prime }e^{-iE_{n+1}^{(2)}t}\left\vert n+1\right\rangle _{2},
\end{eqnarray*}%
we can get the photon states in the lower level%
\begin{eqnarray*}
&&\left\vert \psi (t),\downarrow \right\rangle =\left(
c_{n}e^{-iE_{n}^{(1)}t}e_{n}+c_{n}^{\prime }e^{-iE_{n}^{(2)}t}e_{n}^{\prime
}\right) \left\vert n+1\right\rangle \\
&&+\left( c_{n}e^{-iE_{n}^{(1)}t}f_{n}+f_{n+1}^{\prime
}e^{-iE_{n+1}^{(2)}t}\ e_{n+1}^{\prime }\right) \left\vert n+2\right\rangle .
\end{eqnarray*}%
Then we have the population in the lower state%
\begin{eqnarray*}
P_{\downarrow } &=&c_{n}^{2}(1-c_{n}^{2})+c_{n}^{\prime 2}e_{n}^{\prime
2}+f_{n+1}^{\prime 2}e_{n+1}^{\prime 2} \\
&&+2c_{n}e_{n}c_{n}^{\prime }e_{n}^{\prime }\cos \left[ \left(
E_{n}^{(1)}-E_{n}^{(2)}\right) t\right] \\
&&+2c_{n}f_{n}f_{n+1}^{\prime }e_{n+1}^{\prime }\cos \left[ \left(
E_{n}^{(1)}-E_{n+1}^{(2)}\right) t\right] .
\end{eqnarray*}%
Finally we get qubit population difference $\left\langle \sigma
_{z}\right\rangle =1-2P_{\downarrow }$. Interestingly we obtain two Rabi
frequencies: $\omega _{n}^{(1)}=E_{n}^{(1)}-E_{n}^{(2)}$, $\omega
_{n}^{(2)}=E_{n}^{(1)}-E_{n+1}^{(2)}$, unlike the unmixed model where only
one Rabi frequency in the quantum Rabi oscillation is present. In the
unmixed model with either one- or two-photon terms, $\omega _{n}^{(2)}\ $%
disappears, so the evolution from a number state in the upper level
oscillates sinusoidally.

\begin{figure}[tbp]
\includegraphics[width=8cm]{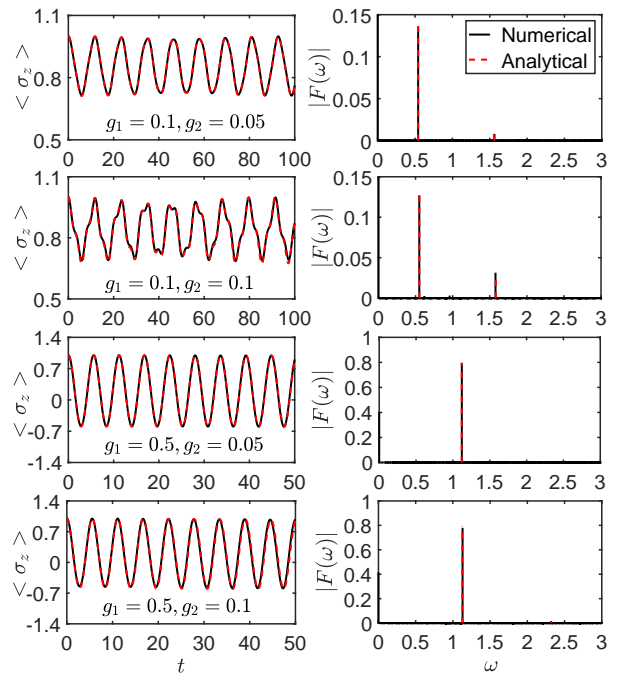}
\caption{(Color online) (left panel ) Population difference at $\Delta=0.5$:
$g_1=0.1, 0.5$, $g_2=0.05, 0.1$ from top to bottom. (right panel ) The
corresponding Fourier transform is calculated. The present analytical
results and numerical ones are indicated by red and black lines. }
\label{RWA_dyn}
\end{figure}

We further examine the dynamics of the population difference $\langle \sigma
_{z}(t)\rangle $ initiated from photonic vacuum states and qubit upper level
at $\Delta =0.5$ for $g_{1}$ in the ultrastrong-coupling regime, and small
values of $g_{2}$. For better understanding, we also analyze the Fourier
transform of $\langle \sigma _{z}(t)\rangle $ in the long time window. The
analytical results are shown in Fig. \ref{RWA_dyn} with red lines. The
numerical exact ones are also collected using black line for comparisons.
The analytical results for $\langle \sigma _{z}(t)\rangle $ match quite well
with the numerical ones. If $g_{1}$ and $g_{2}$ are comparable, two Rabi
frequencies are clearly present with comparable peak height of the Fourier
transform in the upper two panels for $g_{1}=0.1,g_{2}=0.05$ and $0.1$. If $%
g_{1}$ and $g_{2}$ differ considerably, e.g. $g_{1}=0.5$, while $g_{2}=0.05$
and $0.1$, as demonstrated in the lower $2$ panels, Fourier transform
reveals a dominant oscillation with single frequency $\omega _{n}^{(1)}$ for
$g_{1}>>g_{2}$.

Without the nonlinear coupling i.e.$\ g_{2}=0$, it is known that the famous
Rabi oscillation with single frequency occurs. With the presence of the
comparable nonlinear coupling, the two dominant oscillations are present.
With the interplay of the one- and two-photon terms, population difference $%
\langle \sigma _{z}(t)\rangle $ becomes more complicated. If one kind of the
coupling term is relatively weak, and therefore can be omitted, only one of
the two quantum Rabi oscillations is visible.

Rabi oscillations are usually first measured in the newly-built
superconducting qubit and a harmonic oscillator system \cite{Rabi_osc} to
demonstrate the strong coupling. The present emergent novel Rabi
oscillations with two frequencies might be experimentally a signal of the
two competitive couplings.

\section{Vacuum Rabi splitting}

In this section, we apply the analytical results for both the RWA and
non-RWA cases derived above to the famous phenomenon of the vacuum Rabi
splitting \cite{sanche}. In the Jaynes-Cummings model, when an atom is
pumped into an excited state with the vacuum photon state, $\left\vert \psi
_{0}\right\rangle =\left\vert e\right\rangle \left\vert 0\right\rangle $, it
will decay to the ground state via spontaneous emissions. Note that the
ground state of the atom-field system is just the direct product of the
vacuum field and the ground-state atom. The two lowest excited eigenstates
of the system are equivalently observed spectroscopically as a vacuum Rabi
mode splitting resulting in the two-peak emission spectrum. The vacuum Rabi
splitting was experimentally confirmed in many cavity~\cite{Thompson} and
circuit~\cite{Wallraff} QED systems.

In some proposed schemes to realize the qubit and oscillator coupling
systems, the counter-rotating terms can be strongly suppressed \cite%
{AnistropicDicke,Keeling,Fanheng}. In some devices, the anisotropy even
appears quite naturally, because they are controlled by different input
parameters~\cite{Grimsmo1}. Therefore it is technically feasible to realize
the generalized QRM with and without the RWA.

\begin{figure}[tbp]
\includegraphics[width=8cm]{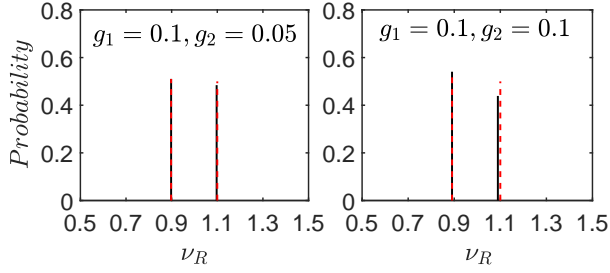}
\caption{ (Color online) Emission spectrum $\protect\nu_R$ in the RWA for
initial state $\left\vert e\right\rangle \left\vert 0\right\rangle $ at $%
g_1=0.1$ for $g_2=0.05$(left) and $g_2=0.1$(right) on resonance ($\Delta =1$%
). The analytical results are indicated by red dashed lines and numerical
exact results by black lines. }
\label{RWA-Rabi-splitting}
\end{figure}

In the present generalized QRM with the RWA, the initial states can be
expanded in terms of states (\ref{wave1}) and (\ref{wave2})
\begin{equation}
\left\vert \psi _{0}\right\rangle =c_{0}\left\vert 0\right\rangle
_{1}+c_{0}^{\prime }\left\vert 0\right\rangle _{2}+f_{1}^{\prime }\left\vert
1\right\rangle _{2}.  \notag
\end{equation}%
The probabilities to find the eigenstates $\left\vert 0\right\rangle _{1}$, $%
\left\vert 0\right\rangle _{2}$, $\left\vert 1\right\rangle _{2}$ in the
initial state are$\ c_{0}^{2},$ $c_{0}^{\prime 2},$and $f_{1}^{\prime 2}$
respectively. The corresponding eigenvalues are$\ E_{0}^{(1)}$, $E_{0}^{(2)}$%
, and $E_{1}^{(2)}$. Then the atom can decay from these three states to the
ground state and the emission spectrum obtained has in principle three peaks
with height proportional to the corresponding probabilities and positions
determined by the corresponding eigenenergies.

In Fig. \ref{RWA-Rabi-splitting}, we plot the analytical emission spectrum
at $g_{1}=0.1$ for $g_{2}=0.05$ and $0.1$ on resonance ($\Delta =1$) with
red dashed lines. Both $g_{1}$ and $g_{2}$ are in the ultrasrong-coupling
regime. The frequency is $\nu _{R}=E-E_{GS}$, where $E_{GS}=-\Delta /2$ is
the ground state energy. Practically, the spectrum has Lorentzian peaks due
to the spontaneous emission to the ground state in the environment. Here the
peaks are shown without width. The numerical results are also presented for
comparison, and good agreement is demonstrated. In principle, three peaks
should appear, but the height ($f_{1}^{\prime 2}$) of the third peak is too
small and omitted here. In the presence of the two-photon terms, the height
of the second peak is lower than that of the first peak, and the peak
difference becomes more pronounced with the two-photon coupling. Note that
in the Jaynes-Cummings model, the emission spectrum has two peaks with equal
probability on resonance. Without the two-photon coupling terms, the effects
of counter-rotating terms on vacuum Rabi splitting have been studied with a
coherent-state approach ~\cite{Zhangyuyu}. The spontaneous emission spectrum
has multiple peaks, and the number of peaks increases with the coupling
strength, in sharp contrast to vacuum Rabi splitting under the RWA.

\begin{figure}[tbp]
\includegraphics[width=8cm]{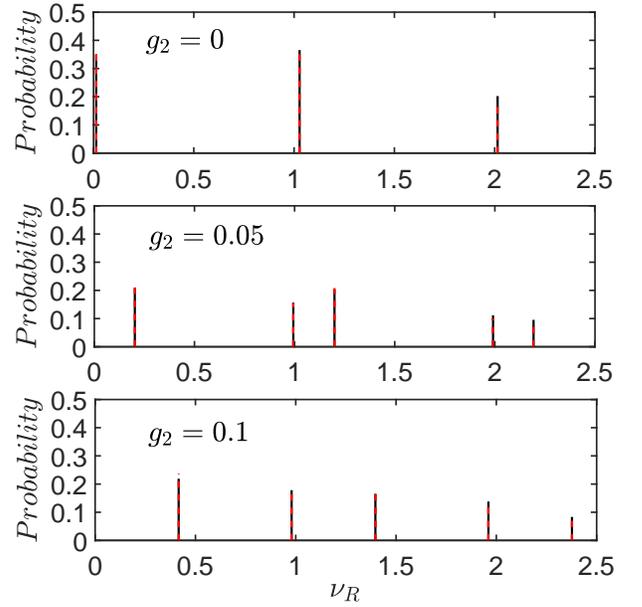}
\caption{ (Color online) Emission spectrum $\protect\nu _{R}$ without the
RWA for initial state $\left\vert e\right\rangle \left\vert 0\right\rangle $
at $\Delta =0.1,g_{1}=1$ for $g_{2}=0$ (top), $0.05$(middle), and $g_{2}=0.1$%
(bottom). The analytical results are indicated by red dashed lines and
numerical exact ones by black lines.}
\label{Full-Rabi-splitting}
\end{figure}
Now we study the effects of two-photon coupling on the vacuum Rabi splitting
without the RWA. In this case, we will employ the analytical results (\ref%
{Full_en}) and (\ref{Full_w}) in the adiabatic approximation. In this case,
we need to transform the wavefunction (\ref{Full_w}) to the original frame
in the representation of $\sigma _{x}$. First, we write the normalized
orthogonal wavefunction in the adiabatic approximation%
\begin{equation*}
\left\vert m\right\rangle ^{\pm }=\left(
\begin{array}{l}
c_{m}^{\pm }\left\vert m\right\rangle _{A} \\
d_{m}^{\pm }\left\vert m\right\rangle _{B}%
\end{array}%
\right) ,
\end{equation*}%
transforming back to the original frame, we have
\begin{equation*}
\left\vert \varphi _{m}\right\rangle ^{\pm }=\frac{1}{\sqrt{2}}\left(
\begin{array}{l}
c_{m}^{\pm }\left\vert m\right\rangle _{A}-d_{m}^{\pm }\left\vert
m\right\rangle _{B} \\
c_{m}^{\pm }\left\vert m\right\rangle _{A}+d_{m}^{\pm }\left\vert
m\right\rangle _{B}%
\end{array}%
\right) .
\end{equation*}%
According to these wavefunctions, the initial state can then be expanded as
\begin{eqnarray*}
\left\vert \psi _{0}\right\rangle &=&\frac{1}{\sqrt{2}}\sum_{m=0}\left\vert
\varphi _{m}\right\rangle ^{-}\left(
c_{m}^{-}D_{m}^{A}-d_{m}^{-}D_{m}^{B}\right) \\
&&+\frac{1}{\sqrt{2}}\sum_{m=0}\left\vert \varphi _{m}\right\rangle
^{+}\left( c_{m}^{+}D_{m}^{A}-d_{m}^{+}D_{m}^{B}\right) ,
\end{eqnarray*}%
where
\begin{eqnarray*}
D_{m}^{A} &=&\frac{1}{\sqrt{um!}}\left( \frac{-v}{2u}\right) ^{m/2}\exp
\left( \frac{-w^{2}(u-v)}{2u}\right) \\
&&H_{m}\left( \frac{w(u-v)}{\sqrt{-2uv}}\right) , \\
D_{m}^{B} &=&\frac{1}{\sqrt{um!}}\left( \frac{v}{2u}\right) ^{m/2}\exp
\left( \frac{-w^{\prime 2}(u+v)}{2u}\right) \\
&&H_{m}\left( \frac{w^{\prime }(u+v)}{\sqrt{2uv}}\right) .
\end{eqnarray*}%
The probability to find the eigenstates $\left\vert m\right\rangle ^{\pm }$
in the initial state is $P_{m}^{\pm }=\frac{1}{2}\left( c_{m}^{\pm
}D_{m}^{A}-d_{m}^{\pm }D_{m}^{B}\right) ^{2}$.

We calculate the emission spectrum at $\Delta =0.1,$ $g_{1}=1.0$ for three
values of $g_{2}$: $0,0.05$, and $0.1$ in Fig. \ref{Full-Rabi-splitting}
both analytically and numerically. The frequency $\nu _{R}=E_{m}^{\pm
}-E_{0}^{-}$ can be obtained with the use of Eq. (\ref{Full_en}). The
analytical results are also in good agreements with the exact numerical
results in this case. Compared to the unmixed one-photon QRM ($g_{2}=0$),
one can find that many more peaks in the spontaneous emission spectrum are
induced when the two-photon coupling term sets in. We have confirmed this
feature extensively with more model parameters.

Theoretically, it is found that some new phenomena emerge with the presence
of the two-photon coupling terms, which may be the possible signal of the
two-photon coupling besides the one-photon coupling in experiments.

\section{ Conclusion}

In summary, the generalized QRM with both one- and two-photon terms is
studied analytically. By using Bogoliubov operators, an effective scheme to
its solution is proposed. The adiabatic approximation, where only the
transition within the same manifold is considered, produces the analytical
eigenvalues and eigenstates in a concise way. Many physical phenomena can
then be \ easily\ analyzed. The obtained energy spectra, which can account
for the experimental transmission spectrum, are in good agreement with the
numerically exact ones in a wide range of coupling strength. The population
dynamics obtained in the adiabatic approximation is also in quantitative
agreement with the numerical ones.

In the RWA, the mathematical simplicity of the eigensolution in the unmixed
Rabi models with either one- or two-photon term is lacking, because of the
absence of the conserved excitation number. We propose an ansatz of the
eigenfunctions including a few dominant Fock states. The corresponding
analytical eigensolutions yield quite good energy levels compared with the
numerical exact ones. With these eigensolutions, we could revisit many
physical problems that have been studied in the unmixed QRM in the RWA. We
study Rabi oscillations here. It is found that the population dynamics can
also match the oscillations for a long time. Two dominant Rabi frequencies
are derived analytically, and further confirmed in numerics.

The concise analytical solution in both full model and in the RWA can be
easily applied in the circuit QED system with the one-photon coupling term
ranging from weak, ultrastrong, to deep-strong-coupling regime for moderate
two-photon coupling. Application of the analytical results to the vacuum
Rabi splitting is performed in this paper as a example. In the RWA, the
different heights of the first two peaks are observed. The second peak
becomes lower with the increase of the two-photon coupling strength. Without
the RWA, more peaks emerge with the advent of the two-photon coupling. These
emergent new phenomena could be detected experimentally if both one- and
two-photon interact with the oscillator simultaneously.

\textbf{ACKNOWLEDGEMENTS} This work is supported by the National Science
Foundation of China (Nos. 11674285, 11834005), the National Key Research and
Development Program of China (No. 2017YFA0303002).

$^{\ast }$ Corresponding author. Email:qhchen@zju.edu.cn \appendix

\section{Solutions in the rotating-wave approximations by the univariate
cubic equation}

In terms of the wavefunction (\ref{wave1}), the Schr$\overset{..}{o}$dinger
equation gives%
\begin{eqnarray}
&&\;\left( n+\frac{\Delta }{2}-E\right) \;c_{n}\left\vert n\right\rangle
+g_{1}e_{n}\sqrt{n+1}\left\vert n\right\rangle  \notag \\
&&+f_{n}\sqrt{n+2}\left\vert n+1\right\rangle  \notag \\
&&+g_{2}\left( e_{n}\sqrt{n\left( n+1\right) }\left\vert n-1\right\rangle
+f_{n}\sqrt{\left( n+2\right) \left( n+1\right) }\left\vert n\right\rangle
\right)  \notag \\
&=&0,  \label{R1}
\end{eqnarray}%
\begin{eqnarray}
&&g_{1}\sqrt{n+1}\left( c_{n}\left\vert n+1\right\rangle \right) +g_{2}\sqrt{%
\left( n+2\right) \left( n+1\right) }c_{n}\left\vert n+2\right\rangle  \notag
\\
&&+\left( n+1-\frac{\Delta }{2}-E\right) e_{n}\left\vert n+1\right\rangle
\notag \\
&&+\left( n+2-\frac{\Delta }{2}-E\right) f_{n}\left\vert n+2\right\rangle
\notag \\
&=&0.  \label{R2}
\end{eqnarray}%
Note that the subspace that wavefunction spanned is not closed, unlike the
unmixed model.

Projecting Eq. (\ref{R1}) onto $\left\vert n\right\rangle $, Eq. (\ref{R2})
onto $\left\vert n+1\right\rangle $ and $\left\vert n+2\right\rangle $, we
have three set equations
\begin{equation*}
\;\left( n+\frac{\Delta }{2}-E\right) \;c_{n}+g_{1}e_{n}\sqrt{n+1},
\end{equation*}%
\begin{equation*}
\;+g_{2}f_{n}\sqrt{\left( n+2\right) \left( n+1\right) }=0,
\end{equation*}%
\begin{eqnarray*}
\;g_{1}\sqrt{n+1}c_{n}+\left( n+1-\frac{\Delta }{2}-E\right) e_{n} &=&0, \\
g_{2}\sqrt{\left( n+2\right) \left( n+1\right) }c_{n}+\left( n+2-\frac{%
\Delta }{2}-E\right) f_{n} &=&0.
\end{eqnarray*}%
Set
\begin{eqnarray*}
x &=&n+1-\frac{\Delta }{2}, \\
y &=&g_{1}\sqrt{n+1}, \\
z &=&g_{2}\sqrt{\left( n+2\right) \left( n+1\right) }.
\end{eqnarray*}%
Nonzero coefficients yield a univariate cubic equation
\begin{equation}
E^{3}+bE^{2}+cE+d=0,  \label{cubic1}
\end{equation}%
where
\begin{eqnarray*}
b &=&-3x-\Delta , \\
c &=&3x^{2}+2\Delta x-y^{2}-z^{2}+\Delta -1, \\
d &=&x-x\Delta +xy^{2}+xz^{2}-x^{2}\Delta -x^{3}+y^{2}.
\end{eqnarray*}

Similarly by the other form of wavefunction Eq. (\ref{wave2}), we can also
obtain three set equations as follows%
\begin{eqnarray*}
\left( n-1+\frac{\Delta }{2}-E\right) \;f_{n}^{\prime }+g_{2}\sqrt{\left(
n+1\right) n}e_{n}^{\prime } &=&0, \\
\left( n+\frac{\Delta }{2}-E\right) \;c_{n}^{\prime }+g_{1}\sqrt{n+1}%
e_{n}^{\prime } &=&0,
\end{eqnarray*}%
\begin{equation*}
g_{2}\sqrt{\left( n+1\right) n}f_{n}^{\prime }+g_{1}\sqrt{n+1}c_{n}^{\prime
}+\left( n+1-\frac{\Delta }{2}-E\right) e_{n}^{\prime }=0.
\end{equation*}%
set%
\begin{eqnarray*}
x^{\prime } &=&n+\frac{\Delta }{2}, \\
y^{\prime } &=&g_{1}\sqrt{n+1}, \\
z^{\prime } &=&g_{2}\sqrt{\left( n+1\right) n},
\end{eqnarray*}%
we can obtain another cubic equation

\begin{equation}
E^{3}+b^{\prime }E^{2}+c^{\prime }E+d^{\prime }=0,  \notag
\end{equation}%
where%
\begin{eqnarray*}
b^{\prime } &=&\Delta -3x^{\prime }, \\
c^{\prime } &=&-y^{\prime 2}-z^{\prime 2}-1+3x^{\prime 2}+\Delta -2x^{\prime
}\Delta , \\
d^{\prime } &=&-x^{\prime 3}-y^{\prime 2}+x^{\prime }\left( 1+y^{\prime
2}+z^{\prime 2}-\Delta \right) +x^{\prime 2}\Delta ,
\end{eqnarray*}

The solutions of the univariate cubic equation
\begin{equation*}
\lambda ^{3}+b\lambda ^{2}+c\lambda +d=0,
\end{equation*}%
can be found in any Mathematics manual. If
\begin{equation*}
\Gamma =B^{2}-4AC<0,
\end{equation*}%
with
\begin{equation*}
A=b^{2}-3c,B=bc-9d,C=c^{2}-3bd,
\end{equation*}%
there are three different real roots
\begin{eqnarray}
\lambda _{1} &=&\;\;\;\;\;\;\frac{-b-2\sqrt{A}\cos \theta }{3},  \label{y1}
\\
\lambda _{2} &=&\frac{-b+\sqrt{A}\left[ \cos \theta -\sqrt{3}\sin \theta %
\right] }{3},  \label{y2} \\
\lambda _{3} &=&\frac{-b+\sqrt{A}\left[ \cos \theta +\sqrt{3}\sin \theta %
\right] }{3},  \label{y3}
\end{eqnarray}%
where
\begin{equation}
\theta =\frac{1}{3}\arccos \left( \frac{2Ab-3B}{2\sqrt{A^{3}}}\right) .
\end{equation}


\end{document}